\documentclass[10pt]{amsart}

\usepackage[all,2cell]{xy} \UseAllTwocells \SilentMatrices
\usepackage{graphicx,amssymb,amsfonts,amsmath,amsthm,newlfont}
\usepackage{epsf}
\usepackage{epsfig}

\usepackage{axodraw4j}
\usepackage{pstricks}

\theoremstyle{definition}

\title{Field diffeomorphisms and the algebraic structure of perturbative expansions}

\author{Dirk Kreimer${}^\ast$
\and 
Andrea Velenich${}^{\ast\ast}$}
\thanks{${}^{\ast}$Alexander von Humboldt Chair in Mathematical Physics supported by the Alexander von Humboldt Foundation and the BMBF. DK also thanks Francis Brown and the IHES for hospitality during a visit to Paris  where
part of this paper was written. Partial support by Brown's ERC grant 257638 is gratefully acknowledged.  ${}^{\ast\ast}$ This work was supported by NSF grant DMS-0603781 in its early stages.}
\address{Depts.\ of Physics and of Mathematics, Humboldt U.\\ Unter den Linden 6\\ 10099 Berlin\\ Germany \and 
Department of Physics,
   Massachusetts Institute of Technology\\
   77 Massachusetts Avenue\\ 
   Cambridge, MA 02139.}
\begin{document}
\maketitle

\bibliographystyle{plain}
\bibliography{main}

\newtheorem{exa}[thm]{Example}

\theoremstyle{definition}

\def\overlap{\;\raisebox{-12mm}{\epsfysize=36mm\epsfbox{overlap.eps}}\;}
\def\decompwthree{\;\raisebox{-8mm}{\epsfysize=24mm\epsfbox{decompwthree.eps}}\;}

\def\oldc{\;\raisebox{-10mm}{\epsfysize=24mm\epsfbox{oldc.eps}}\;}
\def\noldcl{\;\raisebox{-10mm}{\epsfysize=24mm\epsfbox{noldcl.eps}}\;}
\def\noldcr{\;\raisebox{-10mm}{\epsfysize=24mm\epsfbox{noldcr.eps}}\;}
\def\dcl{\;\raisebox{-10mm}{\epsfysize=24mm\epsfbox{dcl.eps}}\;}
\def\dcr{\;\raisebox{-10mm}{\epsfysize=24mm\epsfbox{dcr.eps}}\;}
\def\dclb{\;\raisebox{-10mm}{\epsfysize=24mm\epsfbox{dclb.eps}}\;}
\def\dcrb{\;\raisebox{-10mm}{\epsfysize=24mm\epsfbox{dcrb.eps}}\;}
\def\oot{\;\raisebox{-10mm}{\epsfysize=24mm\epsfbox{oot.eps}}\;}
\def\otf{\;\raisebox{-10mm}{\epsfysize=24mm\epsfbox{otf.eps}}\;}
\def\ofs{\;\raisebox{-10mm}{\epsfysize=24mm\epsfbox{ofs.eps}}\;}
\def\foroldcl{\;\raisebox{-20mm}{\epsfysize=48mm\epsfbox{foroldcl.eps}}\;}
\def\foroldcr{\;\raisebox{-20mm}{\epsfysize=48mm\epsfbox{foroldcr.eps}}\;}

\def\tfour{\;\raisebox{-20mm}{\epsfysize=40mm\epsfbox{tfour.eps}}\;}
\def\tthree{\;\raisebox{-20mm}{\epsfysize=40mm\epsfbox{tthree.eps}}\;}

\def\loglog{\;\raisebox{-12mm}{\epsfysize=24mm\epsfbox{loglog.eps}}\;}

\def\soldc{\;\raisebox{-5mm}{\epsfysize=10mm\epsfbox{oldc.eps}}\;}
\def\soldcbm{\;\raisebox{-5mm}{\epsfysize=10mm\epsfbox{soldcbm.eps}}\;}
\def\soldcbl{\;\raisebox{-5mm}{\epsfysize=10mm\epsfbox{soldcbl.eps}}\;}
\def\soldcbr{\;\raisebox{-5mm}{\epsfysize=10mm\epsfbox{soldcbr.eps}}\;}
\def\soldcbml{\;\raisebox{-5mm}{\epsfysize=10mm\epsfbox{soldcbml.eps}}\;}
\def\soldcbmr{\;\raisebox{-5mm}{\epsfysize=10mm\epsfbox{soldcbmr.eps}}\;}

\def\wttf{\;\raisebox{-10mm}{\epsfysize=20mm\epsfbox{w334.eps}}\;}
\def\wttft{\;\raisebox{-10mm}{\epsfysize=20mm\epsfbox{w3342.eps}}\;}
\def\wtttf{\;\raisebox{-10mm}{\epsfysize=20mm\epsfbox{w3324.eps}}\;}
\def\wttt{\;\raisebox{-10mm}{\epsfysize=20mm\epsfbox{w332.eps}}\;}

\def\wfconn{\;\raisebox{-10mm}{\epsfysize=20mm\epsfbox{wfconn.eps}}\;}
\def\wfconnos{\;\raisebox{-10mm}{\epsfysize=20mm\epsfbox{wfconnos.eps}}\;}

\def\wtwf{\;\raisebox{-15mm}{\epsfysize=36mm\epsfbox{wtwf.eps}}\;}
\def\wtwft{\;\raisebox{-15mm}{\epsfysize=36mm\epsfbox{wtwft.eps}}\;}
\def\wtwfb{\;\raisebox{-15mm}{\epsfysize=36mm\epsfbox{wtwfb.eps}}\;}
\def\wtwftb{\;\raisebox{-15mm}{\epsfysize=36mm\epsfbox{wtwftb.eps}}\;}
\def\wt{\;\raisebox{-10mm}{\epsfysize=24mm\epsfbox{wt.eps}}\;}
\def\wtb{\;\raisebox{-10mm}{\epsfysize=24mm\epsfbox{wtb.eps}}\;}
\def\wtbbull{\;\raisebox{-10mm}{\epsfysize=24mm\epsfbox{wtbbull.eps}}\;}
\def\wtbups{\;\raisebox{-10mm}{\epsfysize=24mm\epsfbox{wtbups.eps}}\;}
\def\wf{\;\raisebox{-10mm}{\epsfysize=24mm\epsfbox{wf.eps}}\;}
\def\wfb{\;\raisebox{-10mm}{\epsfysize=24mm\epsfbox{wfb.eps}}\;}

\def\dcqs{\;\raisebox{-10mm}{\epsfysize=24mm\epsfbox{dcqs.eps}}\;}
\def\dcqssq{\;\raisebox{-10mm}{\epsfysize=24mm\epsfbox{dcqssq.eps}}\;}
\def\subqul{\;\raisebox{-10mm}{\epsfysize=24mm\epsfbox{subqu1.eps}}\;}
\def\codc{\;\raisebox{-10mm}{\epsfysize=24mm\epsfbox{codc.eps}}\;}
\def\codcsq{\;\raisebox{-10mm}{\epsfysize=24mm\epsfbox{codcsq.eps}}\;}

\def\wttfsmall{\;\raisebox{-1mm}{\epsfysize=3mm\epsfbox{w334.eps}}\;}
\def\wttftsmall{\;\raisebox{-1mm}{\epsfysize=3mm\epsfbox{w3342.eps}}\;}
\def\wtttfsmall{\;\raisebox{-1mm}{\epsfysize=3mm\epsfbox{w3324.eps}}\;}
\def\wtttssmall{\;\raisebox{-1mm}{\epsfysize=3mm\epsfbox{w332.eps}}\;}

\def\tltp{\;\raisebox{-10mm}{\epsfysize=24mm\epsfbox{tltp.eps}}\;}
\def\tltpdt{\;\raisebox{-10mm}{\epsfysize=24mm\epsfbox{tltpdt.eps}}\;}
\def\tltpsq{\;\raisebox{-10mm}{\epsfysize=24mm\epsfbox{tltpsq.eps}}\;}
\def\tltpsqdt{\;\raisebox{-10mm}{\epsfysize=24mm\epsfbox{tltpsqdt.eps}}\;}
\def\oltpdot{\;\raisebox{-10mm}{\epsfysize=24mm\epsfbox{oltpdot.eps}}\;}
\def\oltpsq{\;\raisebox{-10mm}{\epsfysize=24mm\epsfbox{oltpsq.eps}}\;}
\def\suboltp{\;\raisebox{-10mm}{\epsfysize=24mm\epsfbox{suboltp.eps}}\;}
\def\suboltpdt{\;\raisebox{-10mm}{\epsfysize=24mm\epsfbox{suboltpdt.eps}}\;}

\def\wts{\;\raisebox{-5mm}{\epsfysize=10mm\epsfbox{wts.eps}}\;}
\def\wtts{\;\raisebox{-5mm}{\epsfysize=10mm\epsfbox{wtts.eps}}\;}
\def\subwts{\;\raisebox{-5mm}{\epsfysize=10mm\epsfbox{subwts.eps}}\;}
\def\subwtts{\;\raisebox{-5mm}{\epsfysize=10mm\epsfbox{subwtts.eps}}\;}
\def\wtssmall{\;\raisebox{-1mm}{\epsfysize=2mm\epsfbox{wts.eps}}\;}
\def\wttssmall{\;\raisebox{-1mm}{\epsfysize=2mm\epsfbox{wtts.eps}}\;}
\def\subwtssmall{\;\raisebox{-1mm}{\epsfysize=2mm\epsfbox{subwts.eps}}\;}
\def\subwttssmall{\;\raisebox{-1mm}{\epsfysize=2mm\epsfbox{subwtts.eps}}\;}

\def\olh{\;\raisebox{-0mm}{\epsfysize=2mm\epsfbox{olh.eps}}\;}
\def\olholh{\;\raisebox{-0mm}{\epsfysize=2mm\epsfbox{olholh.eps}}\;}
\def\olv{\;\raisebox{-1mm}{\epsfysize=4mm\epsfbox{olv.eps}}\;}
\def\tlr{\;\raisebox{-1mm}{\epsfysize=4mm\epsfbox{tlr.eps}}\;}
\def\tll{\;\raisebox{-1mm}{\epsfysize=4mm\epsfbox{tll.eps}}\;}
\def\tlld{\;\raisebox{-8mm}{\epsfysize=20mm\epsfbox{tlld.eps}}\;}
\def\tlg{\;\raisebox{-1mm}{\epsfysize=4mm\epsfbox{tlg.eps}}\;}
\def\flg{\;\raisebox{-1mm}{\epsfysize=4mm\epsfbox{flg.eps}}\;}
\def\tlgo{\;\raisebox{-1mm}{\epsfysize=4mm\epsfbox{tlgo.eps}}\;}
\def\tlgofnr{\;\raisebox{-2mm}{\epsfysize=6mm\epsfbox{tlgofnr.eps}}\;}
\def\tlgofnl{\;\raisebox{-2mm}{\epsfysize=6mm\epsfbox{tlgofnl.eps}}\;}

\def\One{\mathbb{I}}

\begin{abstract}
We consider field diffeomorphisms in the context of real scalar field theories. Starting from free field theories we apply non-linear field diffeomorphisms to the fields and study the perturbative expansion for the transformed theories. We find that tree level amplitudes for the transformed fields must satisfy BCFW type recursion relations for the S-matrix to
remain trivial. 
For the massless field theory these relations continue to hold in loop computations. In the massive field theory the situation is more subtle. A necessary condition for the Feynman rules to respect the maximal ideal and co-ideal defined by the core Hopf algebra of the transformed theory is that upon renormalization all massive tadpole integrals (defined as all integrals independent of the kinematics of external momenta) are mapped to zero.
\end{abstract}

\section{Introduction: From field diffeomorphisms to the core Hopf algebra of graphs}
It has long been known that loop contributions to quantum S-matrix elements can be obtained from tree-level amplitudes using unitarity methods based on the optical theorem and dispersion relations. More recently, these perturbative methods have been applied intensively in QCD  and quantum gravity, where the KLT relations relate the tree-level amplitudes in quantum gravity to the tree-level amplitudes in gauge field theories (\cite{Bern} and references therein). Importantly, $d$-dimensional unitarity methods allow to compute S-matrix elements without the need of an underlying Lagrangean and represent an alternative to the usual quantization prescriptions based on path integrals or canonical quantization. 

In this short paper, we study field diffeomorphisms of a free field theory, which generate a seemingly interacting field theory. 
This is an old albeit somewhat controversial topic in the literature \cite{ACO,castro,KT,nakai,SS,SH,SHL}. We address it here from a minimalistic approach: ignoring any path-integral heuristics, we collect basic facts about the Hopf algebra of a perturbation theory which stems from a field diffeomorphism of a free theory.

As any interacting field theory, an interacting  field theory whose interactions originate from field diffeomorphisms
of a free field theory alone has a perturbative expansion which is governed by a corresponding tower of Hopf algebras \cite{birthday,Suj}. 
It starts from the core Hopf algebra, for which only one-loop graphs are primitive:
\begin{equation}
\Delta \Gamma=\Gamma\otimes\One+\One\otimes\Gamma +\sum_{\cup_i\gamma_i=\gamma\subset\Gamma}\gamma\otimes\Gamma/\gamma,
\end{equation}
 and ends with a Hopf algebra for which any 1-PI graph is primitive:
\begin{equation}
\Delta \Gamma=\Gamma\otimes\One+\One\otimes\Gamma.
\end{equation}
Here, subgraphs $\gamma_i$ are one-particle irreducible (1PI). Intermediate between these two Hopf algebras are those for which graphs of a prescribed superficial degree of divergence contribute
in the coproduct, allowing to treat renormalization and operator product expansions.

All these Hopf algebras allow for maximal co-ideals. In particular, the core Hopf algebra has a maximal  ideal which relates to the celebrated BCFW relations: if the latter relations hold, the Feynman rules are well defined on the quotient of the core Hopf algebra by  this maximal ideal \cite{Suj}.

Gravity as a theory for which the renormalization Hopf algebra equals its core Hopf algebra is a particularly interesting theory from this viewpoint \cite{KrGrav2}. The work here is to be regarded as preparatory work in understanding the algebraic structure of gravity as a quantum field theory. 

\section{Symmetries and Hopf ideals}

In the Hopf algebra of Feynman diagrams, Hopf ideals are known to encode the symmetries of a field  theory \cite{Suj, anatomy}. Such (co-)ideals enforce relations among the n-point 1-particle irreducible Green functions ($\Gamma^{(n)}_{1PI}$) or among the connected Green functions ($\Gamma^{(n)}$), which generically are of the form:
\begin{equation} \label{decore}
\Gamma^{(n)}_{1PI} = \Gamma^{(j)}_{1PI} \frac{1}{\Gamma^{(2)}} \Gamma^{(k)}_{1PI} \qquad \forall \; j,k>2 \; ; \; j+k = n+2.
\end{equation}
and hence
\begin{equation} 
\Gamma^{(n)} = \Gamma^{(j)} \frac{1}{\Gamma^{(2)}} \Gamma^{(k)} \qquad \forall \; j,k>2 \; ; \; j+k = n+2.
\end{equation}
upon iteration.
Here, we use a rather condensed notation where the subscript $j$ indicates  $j$ external fields of some type.

Note that the two-point function is never vanishing: a free field theory provides the lowest order in the perturbation expansion of a field theory, $\Gamma^{(2)}\not=0$ even for vanishing interactions.  
Hence if
$\Gamma^{(3)}=0$ in Eq.(\ref{decore}) we conclude $\Gamma^{(n)}=0,n\geq 3$.  

As an example, relations (\ref{decore}) underlie the BCFW recursive formulae for the computation of tree-level maximally helicity violating (MHV) amplitudes in 
Quantum Chromodynamics \cite{BCFW1,BCFW2}. 
More generally, the relations (\ref{decore}) also hold for the corresponding 1-loop and multi-loop amplitudes in QED and QCD, embodying the gauge symmetry of those theories,
with Ward--Slavnov--Taylor identities being particular instances of such relations when specifying the kinematics of longitudinal and transversal propagation modes.

It is a general graph theoretic result that the sum over 1-particle reducible (1-PR) diagrams can be written in terms of 1-particle irreducible (1-PI) diagrams connected by one or several internal propagators $1/\Gamma^{(2)}$.

Connected $n$-point Feynman diagrams $\Gamma^{(n)}$ are either 1-PR or 1-PI diagrams, but for the massless theory we will show that the 1-loop connected amplitudes vanish when the external legs are evaluated on-shell.
\begin{equation} \label{prima}
\Gamma^{(n)} = \Gamma_{1PI}^{(n)} + \Gamma_{1PR}^{(n)} \qquad \rightarrow \qquad \Gamma_{1PI}^{(n)} = - \Gamma_{1PR}^{(n)}.
\end{equation}

From (\ref{prima}) one obtains Eq.(\ref{decore}) which characterize a Hopf ideal \cite{Suj}. In this case, the Hopf ideal is related to the diffeomorphism invariance of the massless theory.

In the massive theory, instead, we will see below that connected n-point amplitudes do not vanish due to the appearance of tadpole diagrams which spoil the Hopf ideal structure. A necessary condition to regain diffeomorphism invariance is the use of a renormalization scheme 
which eliminates all contributions from tadpole diagrams. This is also a mathematically preferred scheme (see \cite{BrownKr}).
\section{Definitions}
Let us consider real scalar fields defined on a 4-dimensional Minkowski space-time 
$\phi\equiv\phi(x):\mathbb{R}^{1,3} \rightarrow \mathbb{R}$ and field diffeomorphisms $F(\phi)$ specified by choosing a set of real coefficients $\{ a_k \}_{k \in \mathbb{N}}$ which do not depend on the space-time coordinates:
\begin{equation} \label{defF}
F(\phi) = \sum_{k=0}^{\infty} a_k\phi^{k+1} = \phi + a_1 \phi^2 + a_2 \phi^3 + \ldots \qquad (\mathrm{with} \; a_0=1).
\end{equation}
These transformations are called ``point transformations''. They preserve Lagrange's equations, they are a subset of the canonical transformations \cite{castro}, and in the quantum formalism they become unitary transformation of the Hamiltonian \cite{nakai}. 

The two field theories which we will consider are derived from the free massless and the massive scalar field theories, 
with Lagrangean densities $\mathsf{L}[\phi]$ and with $F$ defined as in (\ref{defF}):
\begin{eqnarray}
\label{L1} & \mathsf{L}[\phi]=\frac{1}{2}\partial_\mu\phi\partial^\mu \phi \to \mathsf{L_F}[\phi] = \frac{1}{2} \partial_{\mu} F(\phi) \, \partial^{\mu} F(\phi), \\
\label{L2} & \mathsf{L}[\phi]=\frac{1}{2}\partial_\mu\phi\partial^\mu \phi-\frac{m^2}{2}\phi^2 \to \mathsf{L_F}[\phi] = \frac{1}{2} \partial_{\mu} F(\phi) \, \partial^{\mu} F(\phi) -
 \frac{m^2}{2} F(\phi(x)) F(\phi(x)).
\end{eqnarray}
\section{The massless theory}
Expanding the massless Lagrangean (\ref{L1}) in terms of the field $\phi$, we obtain:
\begin{equation} \label{Lmassless}
\mathsf{L}[\phi] = \frac{1}{2} \partial_{\mu}\phi \partial^{\mu}\phi - \partial_{\mu}\phi \partial^{\mu}\phi \sum_{n=1}^{\infty} \frac{1}{2}\frac{1}{n!} d_n \phi^n,
\end{equation}
where the couplings $d_n$ are defined in terms of the parameters $a_n$ specifying the diffeomorphism $F$:
\begin{equation} \label{ds}
 d_n = n! \sum_{j=0}^n (j+1)(n-j+1) a_j a_{n-j}.
\end{equation}
The Feynman rules for the symmetrized vertices of the Lagrangean (\ref{Lmassless}) are:
\begin{eqnarray} \label{feyn1}
&& \begin{picture}(15,0)\put(-2,2){\includegraphics[width=0.04\textwidth]{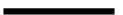}}\end{picture} \quad \rightarrow \quad \frac{i}{k^2} \nonumber \\
&& \begin{picture}(15,0)\put(-1,-4){\includegraphics[width=0.038\textwidth]{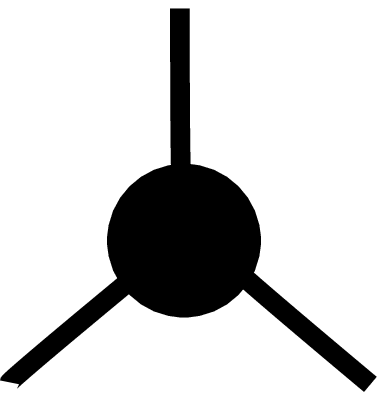}} \end{picture} \quad \rightarrow \quad i \, \frac{d_1}{2}(k_1^2 + k_2^2 + k_3^2) \nonumber \\
&& \begin{picture}(15,0)\put(0,-3){\includegraphics[width=0.035\textwidth]{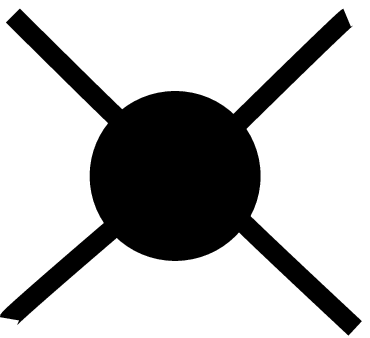}} \end{picture} \quad \rightarrow \quad i \, \frac{d_2}{2}(k_1^2 + \ldots + k_4^2) \\
&& \begin{picture}(15,0)\put(-1,-6){\includegraphics[width=0.04\textwidth]{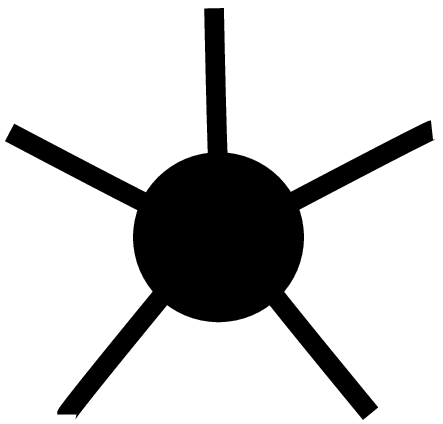}} \end{picture} \quad \rightarrow \quad i \, \frac{d_3}{2}(k_1^2 + \ldots + k_5^2) \nonumber \\
&& \hspace{0.9cm} \ldots \nonumber
\end{eqnarray}
Once evaluated on-shell, the $n$-point tree-level amplitudes vanish for every $n \geq 3$ and, in the classical limit, the field $\phi$ has the same correlations as a free massless scalar field. 

\begin{figure}[!h]
\begin{displaymath}
\begin{picture}(120,45)\put(-10,-3){\includegraphics[width=0.4\textwidth]{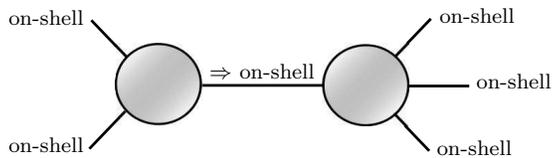}} 
\put(36,25){{\footnotesize $\Rightarrow$ on-shell}} 

\put(-40,45){{\footnotesize on-shell}} 
\put(-40,-3){{\footnotesize on-shell}} 

\put(123,46){{\footnotesize on-shell}} 
\put(137,21){{\footnotesize on-shell}} 
\put(122,-4){{\footnotesize on-shell}} 
\end{picture}
\end{displaymath}
\caption[Example of a tree-level Feynman diagram in which an internal particle becomes physical.]{Example of a tree-level Feynman diagram with an internal propagator (blobs represent arbitrary trees). When all the external legs are on-shell, the internal particle becomes physical. However, instead of contributing a pole to the S-matrix element, the diagram can be recursively shown to vanish. \label{blobs}}
\end{figure}

This is a consequence of the analytic properties of the S-matrix. When a 1-particle intermediate state is physical (i.e.~the internal propagator is on-shell), the S-matrix element is supposed to develop a pole. However, the contribution of the $n$-point vertex to the $n$-point tree-level amplitude vanishes (being proportional to $\sum_{i=1}^n k_i^2$, according to the Feynman rules); any other contribution to the amplitude may only come from tree diagrams with at least one internal propagator (Figure \ref{blobs}). 
An $n$-point tree is partitioned by an internal propagator into two tree-level diagrams, one $m$-point and one $p$-point tree diagrams ($m+p-2=n$). Since the external legs of the $n$-point function are already on-shell, it follows that if the internal propagator is on-shell, then all the external legs of the $m$-point and of the $p$-point amplitudes are on-shell. If the $m$-point and the $p$-point tree-level amplitudes vanish, the corresponding S-matrix element vanishes, just  the opposite of developing  a pole. Since it is easy to show explicitly that the 3-point tree-level amplitudes vanish, the argument above is a recursive proof that all the tree-level S-matrix elements of the theory vanish for $n \geq 3$. 

The same can be obtained by an argument which is by now standard in the study of massless scattering amplitudes.
Assume we have partitioned the $n$-point scattering amplitude $A_n$ as above. Let us shift one of the incoming momenta 
in the amplitude $A_m$ by $q_i\to q_i+zq$, $q^2=0=q_i\cdot q$, for $z$ a complex parameter, and let us shift a momenta of the 
amplitude $A_p$ accordingly, $q_j\to q_j-q z$, $q\cdot q_j=0$.
We obtain a $z$-dependent amplitude $A_n(z)=A_m(z)\frac{1}{(\sum_i p_i+zq)^2}A_p(z)$.
Using our Feynman rules and the fact that $q$ is light-like so that the contour integral in $z$ has no contribution from a residue at infinity, we find that the residue of $A_n(z)/z$ is minus the on-shell residue 
of the intermediate propagator.
We hence find the expected reduction to on-shell evaluations of $A_m,A_p$: we conclude from the fact that $A_3$ vanishes on-shell the vanishing of all higher $n$-point amplitudes. 

\subsection{Loop amplitudes}

The superficial degree of divergence (s.d.d.) of a loop diagram $\Gamma$ computed from the Feynman rules in (\ref{feyn1}) is:
\begin{equation} \label{powcon}
\textrm{s.d.d.}(\Gamma) = |\Gamma|(d-2) + 2,
\end{equation}
where $|\Gamma|$ is the number of loops in the diagram and $d$ is the dimension of space-time. Notably, 
loop diagrams are divergent regardless of the number of their external legs, making the theory non-renormalizable by power counting. 
This is a consequence of the vertices in (\ref{feyn1}) being proportional to the square of the incoming momenta and the propagators 
being proportional to the inverse square of the momentum which they carry, so that the contribution towards the convergence of a loop integration from each propagator is cancelled by the contribution towards divergence from each vertex. A similar power counting appear in the perturbative field theory of gravity \cite{KrGrav}.

The vanishing of the tree-level amplitudes, implies that the 1-loop n-point connected amplitudes vanish for every n$\geq 1$. The proof is a straightforward application of the optical theorem:
\begin{equation} \label{optix}
2 \Im M(\textrm{in} \rightarrow \textrm{out}) = \sum_{\textrm{mid}} \int \prod_{i \in \textrm{mid}} \mathrm{d}^d k_i M^*(\textrm{out} \rightarrow \textrm{mid}) M(\textrm{in} \rightarrow \textrm{mid})
\end{equation}
where ``in'', ``out'' and ``mid'' are the initial, final and intermediate states respectively. 

The optical theorem is equivalent to the application of the Cutkosky rules \cite{cutkosky}: cut the internal propagators in all the possible ways consistent with the fact that the cut legs will be put on shell; replace each cut propagator with a delta function: $(k^2-m^2+i\epsilon)^{-1} \; \rightarrow\ \; -2\pi i \delta(k^2-m^2)$; sum the contributions coming from all possible cuts.

These prescriptions reduce the computation of loop amplitudes to products of on-shell tree-level amplitudes. For the theory described by (\ref{L1}) the vanishing of all tree-level amplitudes implies the vanishing of the 1-loop connected amplitudes,
and similarly at higher loop orders. Note that this implies the use of a kinetic renormalization scheme such that:\\
i)  the finite renormalized amplitudes have the expected 
dispersive properties,\\
 and\\ ii)  the finite renormalized amplitudes do not provide finite parts which are not cut-reconstructible. 

Any minimal subtraction scheme in the context of dimensional regularization would have to be considered problematic in this context, while a kinetic scheme as in \cite{BrownKr} is safe in this respect.
Assuming the use of a kinematic renormalization scheme, no counterterms need then to be added to the Lagrangean and the theory is not only renormalizable,
but indeed respects a trivial maximal co-ideal: 
\begin{equation}
\frac{X^3}{X^2}=\cdots=\frac{X^{n+1}}{X^{n}},
\end{equation}
which is solved by $X^3=0,X^2\not=0$, as expected. Here, the $X^{i}$ are the formal sums over 1PI Feynman graphs with $i$ external legs. 

In the following we will see that the cut-constructibility of loop amplitudes from tree amplitudes does not extend to the massive theory (\ref{L2}) due to 
the appearance of tadpole diagrams which hinder the application of the optical theorem.

\section{The massive theory}

Expanding the massive Lagrangean (\ref{L2}) in terms of the field $\phi$, we obtain:
\begin{equation} \label{Lmassive}
\mathsf{L}[\phi] = \frac{1}{2} \partial_{\mu}\phi \partial^{\mu}\phi - \frac{m^2}{2} \phi^2 + \partial_{\mu}\phi \partial^{\mu}\phi \sum_{n=1}^{\infty} \frac{1}{2} \frac{1}{n!} d_n \phi^n + \sum_{n=1}^{\infty} \frac{1}{(n+2)!} c_n \phi^{n+2}
\end{equation}
where the $\{ d_n \}_{n \in \mathbb{N}}$ are defined as in (\ref{ds}) and the the $\{ c_n \}_{n \in \mathbb{N}}$ are:
\begin{equation} \label{cs}
 c_n = -m^2 \frac{(n+2)!}{2} \sum_{j=0}^n a_j a_{n-j}
\end{equation}
New Feynman rules for the ``massive'' vertices proportional to the couplings $c_n$ complement the Feynman rules in (\ref{feyn1}):
\begin{eqnarray} \label{feyn2}
&& \begin{picture}(15,19)\put(-1,-4){\includegraphics[width=0.038\textwidth]{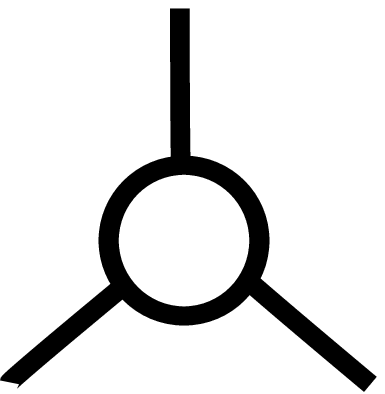}}\end{picture} \quad \rightarrow \quad i \, c_1 \nonumber \\
&& \begin{picture}(15,19)\put(0,-3){\includegraphics[width=0.035\textwidth]{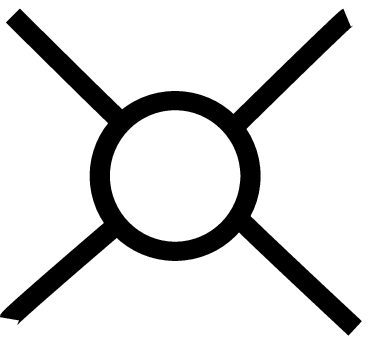}}\end{picture} \quad \rightarrow \quad i \, c_2 \\
&& \begin{picture}(15,19)\put(-1,-6){\includegraphics[width=0.04\textwidth]{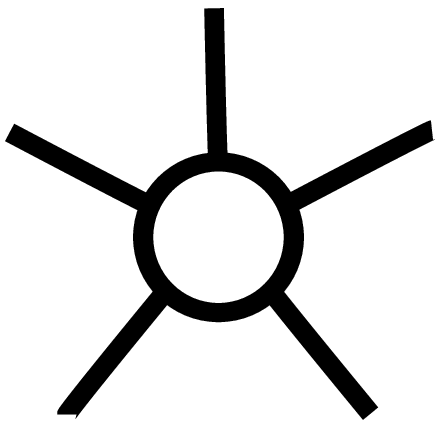}}\end{picture} \quad \rightarrow \quad i \, c_3 \nonumber \\
&& \hspace{0.9cm} \ldots \nonumber
\end{eqnarray}
\subsection{The interplay of propagators and vertices} \label{interplay}
\begin{figure}[!h]
\includegraphics[width=0.5\textwidth]{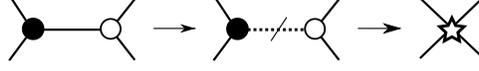}
\caption[Vertices proportional to inverse propagators can effectively modify the topology of Feynman diagrams and generate new contact interactions.]{Vertices proportional to inverse propagators can effectively generate new contact interactions and modify the topology of Feynman diagrams, occasionally creating new tadpoles.\label{fusingfig}}
\end{figure}

Defining the inverse propagators $P_j$'s as:
\begin{displaymath}
\textrm{propagator} \; = \; \frac{i}{P_j} \; = \; \frac{i}{k_j^2 - m^2}
\end{displaymath}
the derivative vertices in (\ref{feyn1}) can be re-written in terms of inverse propagators:
\begin{eqnarray}
&& \begin{picture}(15,0)\put(-1,-4){\includegraphics[width=0.038\textwidth]{3pt0d.eps}} \end{picture} \quad \rightarrow \quad i \, \frac{d_1}{2}(3m^2 + P_1 + P_2 + P_3) \nonumber \\
&& \begin{picture}(15,0)\put(0,-3){\includegraphics[width=0.035\textwidth]{4pt0d.eps}} \end{picture} \quad \rightarrow \quad i \, \frac{d_2}{2}(4m^2 + P_1 + \ldots + P_4) \nonumber \\
&& \begin{picture}(15,0)\put(-1,-6){\includegraphics[width=0.04\textwidth]{5pt0d.eps}} \end{picture} \quad \rightarrow \quad i \, \frac{d_3}{2}(5m^2 + P_1 + \ldots + P_5) \nonumber \\
&& \hspace{0.9cm} \ldots \nonumber
\end{eqnarray}
This formulation clearly shows that vertices with derivatives can then cancel the internal propagator connecting them to a second vertex, effectively fusing with the vertex at the other end of the propagator and generating a new contact interaction (Figure \ref{fusingfig}). These terms typically do not vanish, even when the external legs are on-shell but, surprisingly, explicit computations revealed that after summing all the relevant terms, all the on-shell tree-level amplitudes do vanish, up to the 6-point amplitudes\footnote{These cancellations in general do not happen for any couplings $\{ c_n \}_{n \in \mathbb{N}}$ and $\{ d_n \}_{n \in \mathbb{N}}$ but only for couplings $\{ c_n \}_{n \in \mathbb{N}}$ and $\{ d_n \}_{n \in \mathbb{N}}$ with relations implicitly encoded in (\ref{cs}) and (\ref{ds}).}. A general proof valid for any $n$-point amplitude is, however, still lacking (once again, the explicit checks up to six-point amplitudes can also
be done by promoting internal propagators into complex space, with cancellation of poles at infinity to be explicitly checked upon summing over all contributing massless or massive vertices and over all channels). 

\subsection{Loop amplitudes}

For the massive Lagrangean (\ref{Lmassive}) the divergences of 1-loop Feynman diagrams do not cancel and the n-point 1-loop amplitudes remain divergent. The residues of the 2 and 3-point amplitudes are:
\begin{equation} \label{finalres2}
\textrm{Res}(\textrm{2-pt}) \;\; = \;\; 2 a_1^2 \pi^2 m^2 (2q^2 -m^2) \;\; \stackrel{q^2=m^2}{\rightarrow} \;\; 2 a_1^2 \pi^2 m^4 \nonumber
\end{equation}
\begin{eqnarray} \label{finalres3}
\textrm{Res}(\textrm{3-pt}) &=& (-8 a_1^3 + 12 a_1a_2) \pi^2 m^2 (q_1^2 + q_2^2 + q_3^2) + (24 a_1^3 - 30 a_1a_2) \pi^2 m^4 \nonumber \\
&\stackrel{q^2=m^2}{\rightarrow}& 6 a_1 a_2 \pi^2 m^4 
\end{eqnarray}
Note that, for the 3-point 1-loop amplitude, internal massive and massless vertices of valence 3,4 and 5 contribute so that a rather large class of 
diagrams had to be computed to find the expected reduction to tadpole terms.

The residues of some of the 2- and 3-point Feynman diagrams contain terms proportional to $q^4$, which are not present in the original Lagrangean. However, when the residues relative to all the Feynman diagrams are added up, no terms proportional to $q^4$ are left.
Thus, the derivative terms in the Lagrangean can absorb the $q^2$-dependent part of the residues and the massive terms absorb the $q^2$-independent part. 

Interestingly, the residues of the full 2- and 3-point amplitudes turn out to be proportional to the residue of the corresponding 2- and 3-point tadpoles. Notably, however, the residues of the on-shell 2- and 3-point amplitudes are due not only to tadpole diagrams: they also originate from Feynman diagrams with other topologies whose internal propagators are  cancelled by derivative vertices to generate the before-mentioned contact terms (Figure \ref{fusingfig}). This gives contributions which are effective tadpoles and which renormalize to zero in a kinematic renormalization scheme.
In a kinematic renormalization scheme the subtraction of amplitudes evaluated at different energy scales automatically removes tadpole contributions which, by definition, are independent of the external momenta and thus cancel out in the subtraction.

Partial results on the 4-point massive amplitude hint to the fact that this pattern is likely to extend to generic n-point functions.

Summarizing, we can  end this short first paper  with a conjecture:\\[5mm]
{\em In a kinematic renormalization scheme, massless and massive free field theories are diffeomorphism invariant.}\\[5mm]
Future studies will have to focus on an all order proof of the tree-level recursion and an explicit proof that the non cut-reconstructible amplitudes vanish in kinematic renormalization schemes, as reported here for low orders (in the massive case).

\bibliographystyle{plain}
\renewcommand\refname{References}

\makeatletter
 \def\@biblabel#1{#1}
\makeatother

\end{document}